\documentclass[aip,jap,reprint,showpacs,showkeys,floatfix]{revtex4-1}
\usepackage[pdftex]{graphicx}
\usepackage{bm}
\usepackage{xcolor}
\usepackage{amsmath, amssymb, amsfonts, mathrsfs}
\usepackage{dsfont}
\usepackage{mathrsfs}
\usepackage{setspace}
\usepackage[colorlinks=true]{hyperref}

\renewcommand{\deg}{$^{\circ}$\hspace{1mm}}

\newcommand{\newc}{\newcommand}

\newc{\be}{\begin{equation}}
\newc{\ee}{\end{equation}}
\newc{\bfe}{\begin{floatequation}}
\newc{\efe}{\end{floatequation}}
\newc{\bea}{\begin{eqnarray}}
\newc{\eea}{\end{eqnarray}}
\newc{\ie}{{\it i.e.} }
\newc{\eg}{{\it e.g.} }
\newc{\etc}{{\it etc.} }
\newc{\ra}{\rightarrow}
\newc{\lra}{\leftrightarrow}
\newc{\lsim}{\buildrel\langle\over{\sim}}
\newc{\gsim}{\buildrel\rangle\over{\sim}}

\newc{\one}{\mathds{1}}
\newc{\Tr}[1]{\mathrm{Tr}\left[ {#1} \right]}
\newc{\ket}[1]{\left|{#1}\right\rangle}
\newc{\bra}[1]{\left\langle{#1}\right|}
\newc{\braket}[2]{\langle{#1}|{#2}\rangle}
\newc{\mean}[1]{\langle{#1}\rangle}
\newc{\braketd}[1]{\langle{#1}|{#1}\rangle}
\newc{\ketbrad}[1]{\left|{#1}\rangle\!\langle{#1}\right|}
\newc{\ketbra}[2]{\left|{#1}\rangle\!\langle{#2}\right|}
\newc{\EV}[2]{\langle{#1}\rangle_{#2}}
\newc{\C}{\ensuremath{\mathbbm C}}

\begin{document}

\title{\textcolor{blue}{Multi-user distribution of polarization entangled photon pairs}}

\author{J. Trapateau}
\affiliation{LTCI, CNRS, T\'el\'ecom ParisTech, Universit\'e Paris-Saclay, 75013, Paris, France}
\author{J. Ghalbouni}
\affiliation{Applied Physics Laboratory, Faculty of Sciences 2, Lebanese University, Campus Fanar, BP 90656 Jdedeit, Lebanon}
\author{A. Orieux}
\affiliation{LTCI, CNRS, T\'el\'ecom ParisTech, Universit\'e Paris-Saclay, 75013, Paris, France}
\author{E. Diamanti}
\affiliation{LTCI, CNRS, T\'el\'ecom ParisTech, Universit\'e Paris-Saclay, 75013, Paris, France}
\author{I. Zaquine}\email{isabelle.zaquine@telecom-paristech.fr}
\affiliation{LTCI, CNRS, T\'el\'ecom ParisTech, Universit\'e Paris-Saclay, 75013, Paris, France}
\email{isabelle.zaquine@telecom-paristech.fr}

\begin{abstract}
We experimentally demonstrate multi-user distribution of polarization entanglement using commercial telecom wavelength division demultiplexers. The entangled photon pairs are generated from a broadband source based on spontaneous parametric down conversion in a periodically poled lithium niobate crystal using a double path setup employing a Michelson interferometer and active phase stabilisation. We test and compare demultiplexers based on various technologies and analyze the effect of their characteristics, such as losses and polarization dependence, on the quality of the distributed entanglement for three channel pairs of each demultiplexer. In all cases, we obtain a Bell inequality violation, whose value depends on the demultiplexer features. This demonstrates that entanglement can be distributed to at least three user pairs of a network from a single source. Additionally, we verify for the best demultiplexer that the violation is maintained when the pairs are distributed over a total channel attenuation corresponding to 20 km of optical fiber. These techniques are therefore suitable for resource-efficient practical implementations of entanglement-based quantum key distribution and other quantum communication network applications.
\end{abstract}



\maketitle

\textcolor{blue}{Version: \today}
\section{Introduction}
Quantum entanglement is of paramount importance in quantum communication \cite{GT:natphoton07} and computation \cite{NC:book02}. It is an essential element of quantum information protocols ranging from quantum key distribution \cite{SBC:rmp09} to quantum teleportation \cite{MHS:nature12} and entanglement swapping \cite{HBG:natphys07}, and gives rise to the powerful notion of nonlocality that is tested in Bell inequality experiments \cite{AGR:prl82}.

Entangled photon pairs are typically generated by spontaneous parametric down conversion (SPDC) in nonlinear materials such as bulk crystals \cite{NKZ:apl07,STJ:opex12,CMA:prl13}, semiconductor waveguides \cite{BOA:prl14} and poled fibers \cite{ZTQ:prl12}, or by spontaneous four wave mixing (SFWM) in suitable optical fibers \cite{LCL:opex06,BCT:njp12} and silicon chips \cite{SBO:natphoton14}. Other physical processes for generating entanglement are based, for instance, on quantum dots \cite{HPK:nanolett14,VRJ:natcomm14,TMW:arxiv15}. In the recent years, advances in the performance of entangled photon sources have mainly targeted point-to-point entanglement distribution over long distances \cite{FUH:natphys09,YRL:nature12} while achieving narrowband photon pair generation, compatible with quantum memories, has also been thoroughly investigated \cite{FHP:opex07,BQY:prl08}. These advances are interesting in the context of future communication networks based on quantum repeaters. Another issue of great significance in such large-scale networks is the efficient use of the available resources: indeed, using a source for each pair of users in the network is not a scalable scheme. In this case, the broad bandwidth of photon pairs produced by spontaneous parametric down conversion can allow for entanglement distribution to multiple user pairs from a single source using wavelength division multiplexing techniques. This possibility has been explored in several recent works \cite{LYT:opex08b,ZJD:pra13,HBP:opex13,AM:opex13,ZCG:arxiv15}, while further work is in progress to integrate these devices \cite{MKN:opex14} and to design flexible optical networks based on such sources \cite{CMP:opex14}.

In view of the wide use of wavelength division multiplexing in quantum networks for practical applications, it is essential to be able to properly test the employed demultiplexing technologies and quantify their effect to the quality of the distributed entanglement \cite{GAF:ol13}. In this work, we demonstrate the distribution of polarization entangled photons using three different technologies and provide quality factors that are derived from classical characterization of these devices and that can be used to assess the quality of the obtained quantum correlations. Unlike previous work, where entanglement was characterized using quantum tomography \cite{LYT:opex08b,HBP:opex13} or visibility measurements \cite{AM:opex13}, here we perform Bell tests, in particular we measure in each case the violation of the CHSH inequality \cite{CHSH}. We also perform such tests for a channel attenuation corresponding to 20 km of optical fiber, hence showing that our setup can be readily used, for instance, in multi-party quantum key distribution metropolitan area networks \cite{ZCG:arxiv15}.

\begin{figure}[htbp]
\centering
\includegraphics[angle=0,width=85mm]{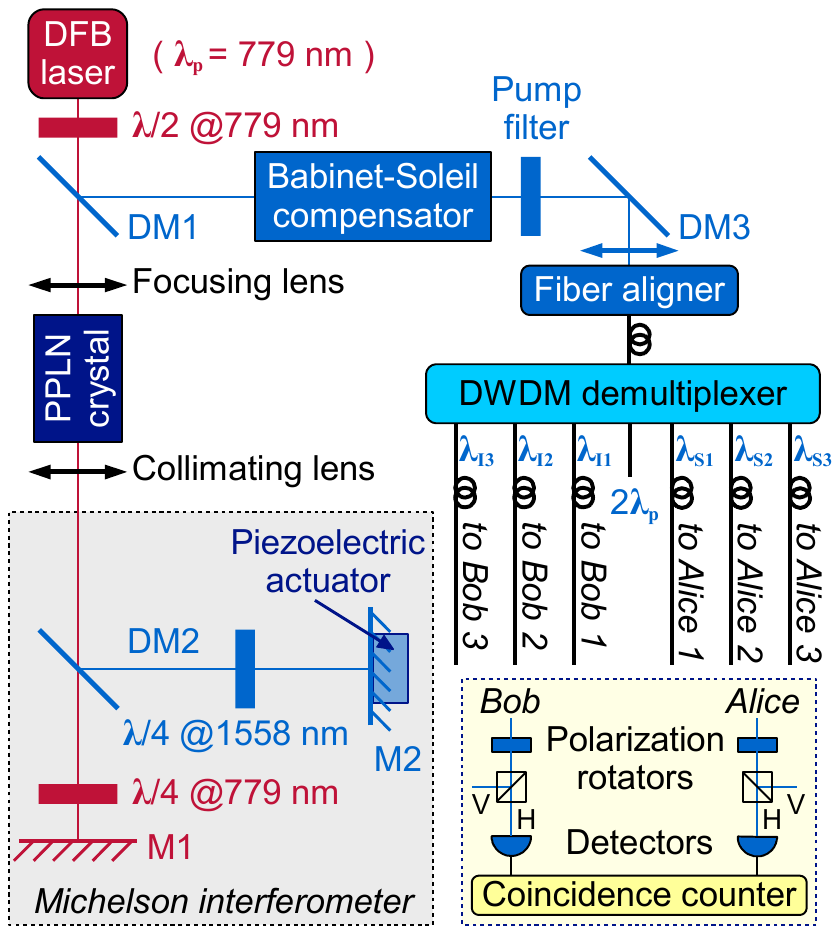}
\caption{(Color online) Experimental setup of the wavelength division multiplexed polarization entangled photon source. (
DFB = Distributed Feedback Laser; DM1, DM2, DM3: dichroic mirrors with $R=99\%$ at 1558 nm and $T=99\%$ at 779 nm; M1,M2: $R=99\%$ at both wavelengths 779 nm and 1558 nm).}
\label{fig:ghaldispositifintrication}
\end{figure}

\section{Experiment}
The experimental setup of our broadband SPDC source of entangled photon pairs is shown in Fig. \ref{fig:ghaldispositifintrication}. The source is based on a double path in a 2 cm long and 500 nm thick LiNbO$_3$ (PPLN) crystal. Light is emitted at 779 nm by a 12 mW continuous wave distributed feedback (DFB) laser. A half wave plate is used first to adjust the proportion of vertical polarisation in the pump beam, which is then focused on the PPLN crystal with a waist of 31 $\mu$m chosen to optimize the coupling of the generated photon pairs into a single mode fiber \cite{SDF:josab13}. The vertical polarisation of the pump produces photon pairs at 1558 nm with a vertical state of polarization, $\ket{VV}$. The role of the dichroic mirror DM2 placed after the crystal is twofold: first, it transmits $99\%$ of the pump beam towards mirror M1 where it is reflected: the double path of the pump beam through a quarter waveplate placed in its path allows the adjustment of the proportion of vertical polarisation of the backward propagating pump (in particular, it can be made larger than the forward propagating one to compensate for losses); second, it reflects the generated photon pairs towards mirror M2 where they are reflected: the double path through a quarter waveplate placed in their path turns their polarisation from vertical to horizontal such that the state $\ket{HH}$ is produced. Next, the backward propagating pump is focused on the crystal leading to the generation of vertically polarized pairs, $\ket{VV}$. The coherent superposition of the two components of the desired entangled state, $\ket{\Psi^+} = (\ket{HH}+\ket{VV}/\sqrt{2}$, is realized on the dichroic mirror DM1. The photon pairs can then be collected in a polarization maintaining fiber after further filtering of the residual pump photons.

It is interesting to note that this entangled photon generation setup is based on a Michelson interferometer, which requires only one nonlinear crystal, unlike the frequently used setups based on Mach Zehnder interferometers. The stability of the interferometer is ensured dynamically using feedback control from the interference of the $1\%$ reflection of the forward propagating pump on DM2, which is reflected by M2 and then transmitted by DM2, with the $1\%$ reflection of the backward propagating pump on DM2 (feedback photodiode not shown in Fig. \ref{fig:ghaldispositifintrication}). This stabilizes the phase difference between the $\ket{VV}$ and the $\ket{HH}$ components of the generated entangled state; once stabilized, this phase difference can be compensated for using a Babinet-Soleil compensator in order to obtain a maximally entangled state.

In our multi-user entanglement distribution setting, the next step is to split the entangled photon pairs using wavelength division demultiplexers. Using the frequency symmetry between the signal and idler photons with respect to half the pump frequency, the entangled photons are coupled to symmetric channels of the demultiplexer. All the devices that we tested had 8 channels within the ITU grid, 100 GHz channel width and 100 GHz channel separation. The photons finally enter the polarization measurement setup, which consists of free space waveplates to choose the measurement basis, fiber polarization beam splitters, and InGaAs single-photon detectors (IDQuantique ID201) featuring a $10\%$ quantum efficiency. 
When triggered at 2 MHz with a $10 \mu$s dead time and 20 ns gates, the detectors had around 500 dark counts per second. A fast time-to-digital converter was used as a time interval analyzer to count the coincidence events between the two paths, with a coincidence gate of 1 ns.

\begin{table*}[htbp]
\begin{center}
\begin{tabular}{ |c || c | c | c | c || c | c | c | }
\hline
Demultiplexer & & & & Brightness & & \\ 
(ITU channel pair) & $V_0$ &$V_{45}$  &$S$  & (true coinc. &$\zeta_{Q}$  &$\eta$  \\ 
& & & & per min) & & \\ \hline \hline
DTF (22,26)  & \,$0.88\pm0.02$\, & \,$0.88\pm0.02$\, & \,$2.40 \pm 0.05 $\, & 547 & \,0.42\, & \,$0.982\pm0.026$\, \\ \hline
DTF (21,27)  & \,$0.85\pm0.02$\, & \,$0.87\pm0.02$\, & \,$2.57 \pm 0.05 $\, & 540 & \,0.49\, & \,$1.000\pm0.028$\, \\ \hline
DGG (23,25)  & \,$0.89\pm0.04$\, & \,$0.76\pm0.04$\, & \,$2.39 \pm 0.08$\, & 176 & \,0.19\, & \,$0.938\pm0.048$\, \\ \hline
DGG (22,26)  & \,$0.80\pm0.04$\, & \,$0.79\pm0.04$\, & \,$2.21 \pm 0.09 $\, & 172 & \,0.19\, & \,$0.985\pm0.056$\, \\ \hline
DGG (21,27)  & \,$0.82\pm0.04$\, & \,$0.77\pm0.04$\, & \,$ 2.25 \pm 0.07$\, & 216 & \,0.20\, & \,$0.970\pm0.053$\, \\ \hline
AWG (23,25)  & \,$0.77 \pm0.04$\, & \,$0.74\pm0.05$\, & \,$2.10 \pm 0.10 $\, & 149 & \,0.064\, & \,$0.972\pm0.059$\, \\ \hline
AWG (22,26)  & \,$0.79 \pm0.05$\, & \,$0.66\pm0.05$\, & \,$2.00 \pm 0.10 $\, & 116 & \,0.079\, & \,$0.906\pm0.066$\, \\ \hline
DGFT (23,25)  & \,$0.79\pm0.05$\, & \,$0.82\pm0.05$\, & \,$2.30 \pm 0.10 $\, & 108 & \,0.030\, & \,$1.000\pm0.072$\, \\ \hline
DGFT (22,26)  & \,$0.80\pm0.05$\, & \,$0.80\pm0.05$\, & \,$2.20 \pm 0.10 $\, & 130 & \,0.035\, & \,$0.986\pm0.069$\, \\ \hline
DGFT (21,27)  & \,$0.81\pm0.05$\, & \,$0.75\pm0.05$\, & \,$2.40  \pm 0.10 $\, & 113 & \,0.035\, & \,$1.000\pm0.069$\, \\ \hline
\end{tabular}
\caption{Visibility $V_0$ in the 0\deg (natural) basis and $V_{45}$ in the 45\deg (diagonal) basis, measured Bell parameter $S$ and brightness. The quality factor $\zeta_{Q}$ and polarization mode overlap coefficient $\eta$ 
 were estimated from $V_0$, $V_{45}$ and $S$. The channel numbers correspond to the International Telecommunication Union (ITU) grid. The uncertainties have been calculated assuming a Poisson distribution of the counts. }
\label{table:table1}
\end{center}
\end{table*}

We performed our tests using four different types of commercial DWDM devices that are based on three main technologies \cite{CC:book03}: (a) dielectric thin-film (DTF), consisting of fiber Fabry-Perot cavities: these transmission bandpass filters are cascaded in order to separate the different channels; (b) arrayed-waveguide gratings (AWG) that are planar lightwave circuits based on multibeam interference; and (c) free space diffraction gratings (DG) used in combination with imaging optics: due to wavelength-dependent diffraction angles, diffracted beams with different wavelengths are focused into different locations and then coupled into output fibers. In all three technologies, the spectral shape of the transmission curves can be either Gaussian or flat-top. The DTF and AWG devices we tested were flat-top, while we tested both a Gaussian DG (DGG) and a flat-top one (DGFT).

In order to evaluate the quality of the distributed entanglement and assess the effect of the various demultiplexers, we measure the following parameters: the visibility in the natural and the diagonal bases, $V = (C_{\text{max}}-C_{\text{min}})/(C_{\text{max}}+C_{\text{min}})$, where $C_{\text{max}}$ and $C_{\text{min}}$ are respectively the maximum and minimum number of coincidences when one of the polarisation basis angles is changed; the violation of the CHSH inequality, which is quantified by the Bell parameter $S$; and the brightness $B$. The results of our measurements are given in Table \ref{table:table1}. In order to be meaningful to a user, the source brightness is defined as the number of \emph{true coincidences} for a given time length, spectral bandwidth and pump power. This means that the number of accidental coincidences, which are not due to correlated events, has been calculated (the probability of accidentals is upper bounded by the product of the probabilities of counts on the two paths\cite{KSC:ao94}) and subtracted from the total coincidence number; hence the reported results correspond to a lower bound of the number of entangled pairs (useful pairs) produced by the source. 

\begin{figure}[htbp]
\includegraphics[width=85mm]{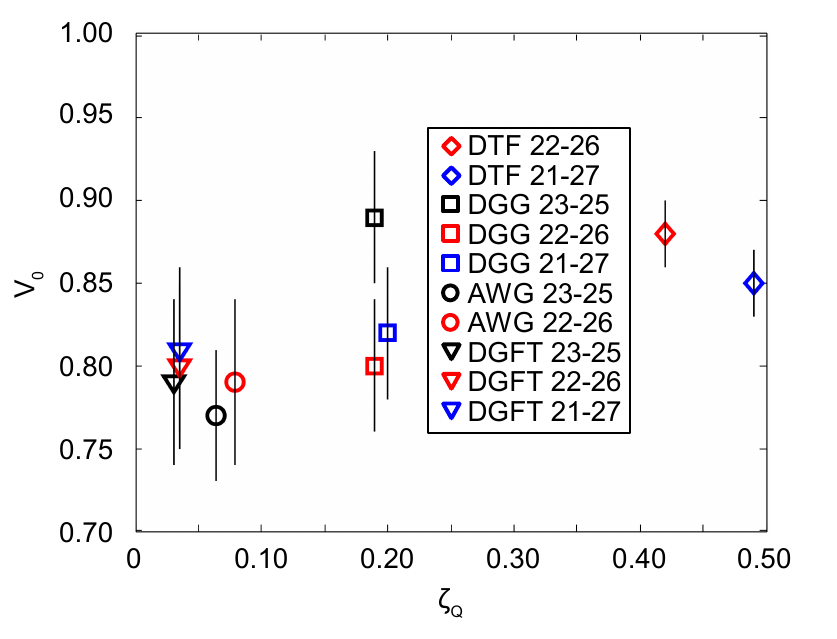}
\caption{(Color online) The visibility $V_0$ is plotted as a function of the quality factor $\zeta_{Q}$.}
\label{fig:V0vsB}
\end{figure}

In Fig. \ref{fig:V0vsB}, we plot, for all tested channel pairs, the measured visibility in the natural basis $V_0$ as a function of the quality factor $\zeta_{Q}$ defined as:
\begin{equation} \label{eq:zetaQ}
\zeta_{Q} = \frac{I_2^2}{I_{1_A}I_{1_B}}T_AT_B
\end{equation}
In this expression, $T_A$ and $T_B$ are the maximum transmission efficiencies of the two demultiplexer output channels, $A$ and $B$ (corresponding to Alice and Bob respectively in Fig. \ref{fig:ghaldispositifintrication}), while $I_{1_i}$ ($i = A, B$) and $I_{2}$ are defined as follows:
\begin{eqnarray}
I_{1_i} &=& \int_{\nu_{i_1}}^{\nu_{i_2}} \frac{d\nu}{2 \pi}\tau(\nu - \nu_{i_{C}})  \nonumber \\
I_{2} (\nu_{ij_C},\nu_p)&=& \int_{\nu_{i_1}}^{\nu-{j_2}}    \frac{d\nu_s}{2 \pi}\tau(\nu_s - \nu_{ij_{C}})\tau(\nu_p-\nu_s - \nu_{ij_{C}}), \nonumber
\end{eqnarray}
where $\tau(\nu)$ is the normalized intensity transmission of the demultiplexer channel $i$, $\nu_{i_{C}}$, $\nu_{i_{1}}$ and $\nu_{i_{2}}$ are respectively the center, start and stop frequencies of channel $i$, and $\nu_{ij_C}$ is the middle frequency corresponding to channel pair $ij$ \cite{GAF:ol13}. Let us note that $I_2$ is maximum when $\nu_{ij_C}=\nu_p/2$. For instance, in our case the frequency $\nu_p/2$ corresponds to the ITU channel 24, which means that we expect to detect quantum correlations in the channel pairs that are symmetric with respect to this channel (see Table \ref{table:table1}). The expression of Eq. (\ref{eq:zetaQ}) is deduced from the rate of the decrease of the maximum visibility that can be expected for a given demultiplexer with respect to the expected brightness (See Appendix ref{appendix:zetaQ}).
This quality factor essentially lumps into a single quantity the effect of the characteristics, in particular the losses and the spectral features, of the employed demultiplexers to the distributed quantum correlations. Its maximum value is $\zeta_{Q}=1$. The values of $\zeta_{Q}$ calculated from the measured transmission $\tau(\nu)$ for all channel pairs used in our experiments are given in Table \ref{table:table1}.

\begin{figure}[htbp]
\includegraphics[width=85mm]{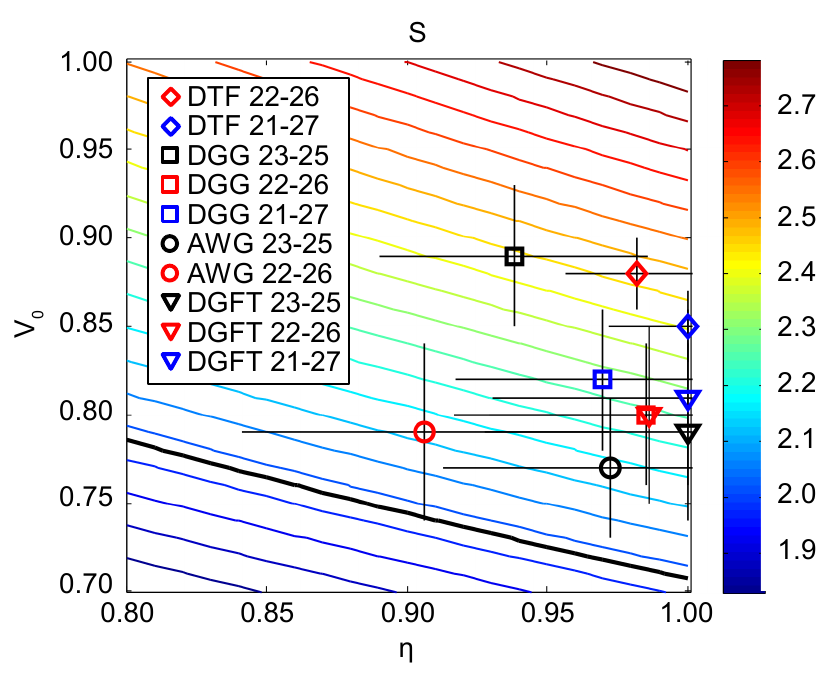}
\caption{(Color online) $S$ is plotted as a function of $V_0$ and $\eta$. The color lines are a contour plot of $S(V_0,\eta)$ as expressed in Eq. (\ref{eq:Svseta}). The symbols represent the measured values for the 10 tested channel pairs.}
\label{fig:SvsV0eteta}
\end{figure}
\section{Discussion}
It is interesting to remark that the obtained values for $\zeta_{Q}$ provide an indicative classification of the demultiplexers; for instance, the dielectric thin film technology (DTF) clearly stands out as the one providing the best performance in terms of brightness and entanglement preservation, due mainly to its small losses. However, this quality factor does not capture the full behavior that we would like to assess. For example, the violation obtained with the arrayed-waveguide grating (AWG) demultiplexers is smaller than the one obtained with the flat-top shaped diffraction gratings (DGFT) despite the higher $\zeta_{Q}$ values of the former. This is due to the fact that in order to perfectly preserve the polarization entanglement, no discernability should be introduced between the $H$ and $V$ polarization modes. This is not the case in the presence of polarization mode dispersion (PMD), which introduces a polarization-dependent delay that can be detrimental to entanglement if it is comparable to the coherence time of the photons. Note that the input state $\ket{\Phi^+}$ is  decoherence free \cite{WJM:pra01} under collective PMD (i.e. when the PMD is the same in both channels): only the difference in PMD between the channel induces decoherence here. Indeed, one can show (See Appendix \ref{appendix:PMD}).
that the Bell parameter $S$ can be expressed in this case as:
\begin{equation}\label{eq:Svseta}
S = \sqrt{2}(V_0+V_{45}) = \sqrt{2}V_0(1+\eta),
\end{equation}
where
\begin{equation}\label{eq:eta}
\eta =  \int f_s(t)f_i(\tau_{\text{PMD}}-t)dt,
\end{equation}
is the temporal overlap between the vertical photon wavepackets of the signal and idler demultiplexer channels, assuming that the horizontal ones overlap perfectly, $\tau_{\text{PMD}}$ is the delay due to the differential PMD between the two channels and $f_s(t)$ and $f_i(t)$ are the normalized temporal transmission functions of the signal and idler channels (for a Fourier limited 100 GHz Gaussian channel for instance, the full width at half-maximum of the transmission function is 4.5 ps).

The PMD values are usually not specified by demultiplexer manufacturers because this is not a critical parameter for classical telecommunication applications. From our measurement results, we could deduce the values of the coefficient $\eta$ for each tested channel pair from the measured values of $V_0$ and $S$ by inverting Eq. (\ref{eq:Svseta}). The results are shown in the contour plot of Fig. \ref{fig:SvsV0eteta}, while the values for the parameters $\eta$ 
are given in Table \ref{table:table1}. The corresponding values of $\tau_{\text{PMD}}$ remain smaller than $2$ ps. The DTF and DGFT technologies present very little to no PMD allowing the entanglement to reach the limit set by $V_0$. The DGG and AWG technologies, however, both introduce a non-negligible amount of PMD that degrades the entanglement.

This additional analysis, quantified by the parameter $\eta$, allows us to assess more precisely the effect of the demultiplexing technology to the entanglement distribution in this DWDM setting.  Indeed, when both quality factors and corresponding figures are considered, we can confirm that the DTF technology gives the best results for both brightness and entanglement preservation, while the DG technology has lower brightness in general but preserves well the entanglement despite the relatively high PMD for the gaussian case. The AWG technology on the other hand is disadvantaged by the high PMD that it features.

In order to test our wavelength multiplexed entanglement source in a realistic communication setting, we performed experiments for the dielectric thin film demultiplexer channel pair (21,27) introducing in the paths of the distributed photons a fixed attenuation corresponding to the transmission along a 10 km optical fiber each, hence to a 20 km total distance between the users. The measured Bell parameter in this case was $S = 2.24 \pm 0.09$. This result illustrates the suitability of our source for practical applications within quantum networks requiring efficient use of the available resources.

\section{Conclusion}
In summary we have demonstrated multi-user distribution of entanglement using a single polarization entangled photon source, stabilized with a Michelson interferometer setup. Two figures of merit have been defined to assess the performance of the four types of demultiplexers based on three different technologies : $\zeta_{Q}$ that can be calculated from classical transmission measurements of the demultiplexer and $\eta$ that can de deduced from the measured visibilities in the 0° and 45° bases and takes into account the polarisation mode dispersion of the demultiplexers. We believe that the resulting classification of these commercial components and the test of entanglement preservation in the realistic situation of channel attenuation can be very useful in the context of quantum networks. 

\section*{Acknowledgments}
This research was supported by Institut T\'el\'ecom - T\'el\'ecom ParisTech and the Ile-de-France region through the project QUIN.


\bibliography{JAP_JT}

\begin{thebibliography}{34}%
\makeatletter
\providecommand \@ifxundefined [1]{%
 \@ifx{#1\undefined}
}%
\providecommand \@ifnum [1]{%
 \ifnum #1\expandafter \@firstoftwo
 \else \expandafter \@secondoftwo
 \fi
}%
\providecommand \@ifx [1]{%
 \ifx #1\expandafter \@firstoftwo
 \else \expandafter \@secondoftwo
 \fi
}%
\providecommand \natexlab [1]{#1}%
\providecommand \enquote  [1]{``#1''}%
\providecommand \bibnamefont  [1]{#1}%
\providecommand \bibfnamefont [1]{#1}%
\providecommand \citenamefont [1]{#1}%
\providecommand \href@noop [0]{\@secondoftwo}%
\providecommand \href [0]{\begingroup \@sanitize@url \@href}%
\providecommand \@href[1]{\@@startlink{#1}\@@href}%
\providecommand \@@href[1]{\endgroup#1\@@endlink}%
\providecommand \@sanitize@url [0]{\catcode `\\12\catcode `\$12\catcode
  `\&12\catcode `\#12\catcode `\^12\catcode `\_12\catcode `\%12\relax}%
\providecommand \@@startlink[1]{}%
\providecommand \@@endlink[0]{}%
\providecommand \url  [0]{\begingroup\@sanitize@url \@url }%
\providecommand \@url [1]{\endgroup\@href {#1}{\urlprefix }}%
\providecommand \urlprefix  [0]{URL }%
\providecommand \Eprint [0]{\href }%
\providecommand \doibase [0]{http://dx.doi.org/}%
\providecommand \selectlanguage [0]{\@gobble}%
\providecommand \bibinfo  [0]{\@secondoftwo}%
\providecommand \bibfield  [0]{\@secondoftwo}%
\providecommand \translation [1]{[#1]}%
\providecommand \BibitemOpen [0]{}%
\providecommand \bibitemStop [0]{}%
\providecommand \bibitemNoStop [0]{.\EOS\space}%
\providecommand \EOS [0]{\spacefactor3000\relax}%
\providecommand \BibitemShut  [1]{\csname bibitem#1\endcsname}%
\let\auto@bib@innerbib\@empty
\bibitem [{\citenamefont {Gisin}\ and\ \citenamefont
  {Thew}(2007)}]{GT:natphoton07}%
  \BibitemOpen
  \bibfield  {author} {\bibinfo {author} {\bibfnamefont {N.}~\bibnamefont
  {Gisin}}\ and\ \bibinfo {author} {\bibfnamefont {R.}~\bibnamefont {Thew}},\
  }\href@noop {} {\bibfield  {journal} {\bibinfo  {journal} {Nature Photon.}\
  }\textbf {\bibinfo {volume} {1}},\ \bibinfo {pages} {165} (\bibinfo {year}
  {2007})}\BibitemShut {NoStop}%
\bibitem [{\citenamefont {Nielsen}\ and\ \citenamefont
  {Chuang}(2002)}]{NC:book02}%
  \BibitemOpen
  \bibfield  {author} {\bibinfo {author} {\bibfnamefont {M.~A.}\ \bibnamefont
  {Nielsen}}\ and\ \bibinfo {author} {\bibfnamefont {I.~L.}\ \bibnamefont
  {Chuang}},\ }\href@noop {} {\emph {\bibinfo {title} {Quantum computation and
  quantum information}}}\ (\bibinfo  {publisher} {Cambridge},\ \bibinfo {year}
  {2002})\BibitemShut {NoStop}%
\bibitem [{\citenamefont {Scarani}\ \emph {et~al.}(2009)\citenamefont
  {Scarani}, \citenamefont {Bechmann-Pasquinucci}, \citenamefont {Cerf},
  \citenamefont {Du{\v{s}}ek}, \citenamefont {L{\"u}tkenhaus},\ and\
  \citenamefont {Peev}}]{SBC:rmp09}%
  \BibitemOpen
  \bibfield  {author} {\bibinfo {author} {\bibfnamefont {V.}~\bibnamefont
  {Scarani}}, \bibinfo {author} {\bibfnamefont {H.}~\bibnamefont
  {Bechmann-Pasquinucci}}, \bibinfo {author} {\bibfnamefont {N.~J.}\
  \bibnamefont {Cerf}}, \bibinfo {author} {\bibfnamefont {M.}~\bibnamefont
  {Du{\v{s}}ek}}, \bibinfo {author} {\bibfnamefont {N.}~\bibnamefont
  {L{\"u}tkenhaus}}, \ and\ \bibinfo {author} {\bibfnamefont {M.}~\bibnamefont
  {Peev}},\ }\href@noop {} {\bibfield  {journal} {\bibinfo  {journal} {Rev.
  Mod. Phys.}\ }\textbf {\bibinfo {volume} {81}},\ \bibinfo {pages} {1301}
  (\bibinfo {year} {2009})}\BibitemShut {NoStop}%
\bibitem [{\citenamefont {Ma}\ \emph {et~al.}(2012)\citenamefont {Ma},
  \citenamefont {Herbst}, \citenamefont {Scheidl}, \citenamefont {Wang},
  \citenamefont {Kropatschek}, \citenamefont {Naylor}, \citenamefont
  {Wittmann}, \citenamefont {Mech}, \citenamefont {Kofler}, \citenamefont
  {Anisimova}, \citenamefont {Makarov}, \citenamefont {Jennewein},
  \citenamefont {Ursin},\ and\ \citenamefont {Zeilinger}}]{MHS:nature12}%
  \BibitemOpen
  \bibfield  {author} {\bibinfo {author} {\bibfnamefont {X.-S.}\ \bibnamefont
  {Ma}}, \bibinfo {author} {\bibfnamefont {T.}~\bibnamefont {Herbst}}, \bibinfo
  {author} {\bibfnamefont {T.}~\bibnamefont {Scheidl}}, \bibinfo {author}
  {\bibfnamefont {D.}~\bibnamefont {Wang}}, \bibinfo {author} {\bibfnamefont
  {S.}~\bibnamefont {Kropatschek}}, \bibinfo {author} {\bibfnamefont
  {W.}~\bibnamefont {Naylor}}, \bibinfo {author} {\bibfnamefont
  {B.}~\bibnamefont {Wittmann}}, \bibinfo {author} {\bibfnamefont
  {A.}~\bibnamefont {Mech}}, \bibinfo {author} {\bibfnamefont {J.}~\bibnamefont
  {Kofler}}, \bibinfo {author} {\bibfnamefont {E.}~\bibnamefont {Anisimova}},
  \bibinfo {author} {\bibfnamefont {V.}~\bibnamefont {Makarov}}, \bibinfo
  {author} {\bibfnamefont {T.}~\bibnamefont {Jennewein}}, \bibinfo {author}
  {\bibfnamefont {R.}~\bibnamefont {Ursin}}, \ and\ \bibinfo {author}
  {\bibfnamefont {A.}~\bibnamefont {Zeilinger}},\ }\href@noop {} {\bibfield
  {journal} {\bibinfo  {journal} {Nature}\ }\textbf {\bibinfo {volume} {489}},\
  \bibinfo {pages} {269} (\bibinfo {year} {2012})}\BibitemShut {NoStop}%
\bibitem [{\citenamefont {Halder}\ \emph {et~al.}(2007)\citenamefont {Halder},
  \citenamefont {Beveratos}, \citenamefont {Gisin}, \citenamefont {Scarani},
  \citenamefont {Simon},\ and\ \citenamefont {Zbinden}}]{HBG:natphys07}%
  \BibitemOpen
  \bibfield  {author} {\bibinfo {author} {\bibfnamefont {M.}~\bibnamefont
  {Halder}}, \bibinfo {author} {\bibfnamefont {A.}~\bibnamefont {Beveratos}},
  \bibinfo {author} {\bibfnamefont {N.}~\bibnamefont {Gisin}}, \bibinfo
  {author} {\bibfnamefont {V.}~\bibnamefont {Scarani}}, \bibinfo {author}
  {\bibfnamefont {C.}~\bibnamefont {Simon}}, \ and\ \bibinfo {author}
  {\bibfnamefont {H.}~\bibnamefont {Zbinden}},\ }\href@noop {} {\bibfield
  {journal} {\bibinfo  {journal} {Nature Phys.}\ }\textbf {\bibinfo {volume}
  {3}},\ \bibinfo {pages} {692} (\bibinfo {year} {2007})}\BibitemShut {NoStop}%
\bibitem [{\citenamefont {Aspect}, \citenamefont {Grangier},\ and\
  \citenamefont {Roger}(1982)}]{AGR:prl82}%
  \BibitemOpen
  \bibfield  {author} {\bibinfo {author} {\bibfnamefont {A.}~\bibnamefont
  {Aspect}}, \bibinfo {author} {\bibfnamefont {P.}~\bibnamefont {Grangier}}, \
  and\ \bibinfo {author} {\bibfnamefont {G.}~\bibnamefont {Roger}},\
  }\href@noop {} {\bibfield  {journal} {\bibinfo  {journal} {Phys. Rev. Lett.}\
  }\textbf {\bibinfo {volume} {49}},\ \bibinfo {pages} {91} (\bibinfo {year}
  {1982})}\BibitemShut {NoStop}%
\bibitem [{\citenamefont {Noh}, \citenamefont {Kim},\ and\ \citenamefont
  {Zyung}(2007)}]{NKZ:apl07}%
  \BibitemOpen
  \bibfield  {author} {\bibinfo {author} {\bibfnamefont {T.-G.}\ \bibnamefont
  {Noh}}, \bibinfo {author} {\bibfnamefont {H.}~\bibnamefont {Kim}}, \ and\
  \bibinfo {author} {\bibfnamefont {T.}~\bibnamefont {Zyung}},\ }\href@noop {}
  {\bibfield  {journal} {\bibinfo  {journal} {App. Phys. Lett.}\ }\textbf
  {\bibinfo {volume} {90}},\ \bibinfo {pages} {011116} (\bibinfo {year}
  {2007})}\BibitemShut {NoStop}%
\bibitem [{\citenamefont {Steinlechner}\ \emph {et~al.}(2012)\citenamefont
  {Steinlechner}, \citenamefont {Trojek}, \citenamefont {Jofre}, \citenamefont
  {Weier}, \citenamefont {Perez}, \citenamefont {Jennewein}, \citenamefont
  {Ursin}, \citenamefont {Rarity}, \citenamefont {Mitchell}, \citenamefont
  {Torres}, \citenamefont {Weinfurter},\ and\ \citenamefont
  {Pruneri}}]{STJ:opex12}%
  \BibitemOpen
  \bibfield  {author} {\bibinfo {author} {\bibfnamefont {F.}~\bibnamefont
  {Steinlechner}}, \bibinfo {author} {\bibfnamefont {P.}~\bibnamefont
  {Trojek}}, \bibinfo {author} {\bibfnamefont {M.}~\bibnamefont {Jofre}},
  \bibinfo {author} {\bibfnamefont {H.}~\bibnamefont {Weier}}, \bibinfo
  {author} {\bibfnamefont {D.}~\bibnamefont {Perez}}, \bibinfo {author}
  {\bibfnamefont {T.}~\bibnamefont {Jennewein}}, \bibinfo {author}
  {\bibfnamefont {R.}~\bibnamefont {Ursin}}, \bibinfo {author} {\bibfnamefont
  {J.}~\bibnamefont {Rarity}}, \bibinfo {author} {\bibfnamefont {M.~W.}\
  \bibnamefont {Mitchell}}, \bibinfo {author} {\bibfnamefont {J.~P.}\
  \bibnamefont {Torres}}, \bibinfo {author} {\bibfnamefont {H.}~\bibnamefont
  {Weinfurter}}, \ and\ \bibinfo {author} {\bibfnamefont {V.}~\bibnamefont
  {Pruneri}},\ }\href@noop {} {\bibfield  {journal} {\bibinfo  {journal} {Opt.
  Express}\ }\textbf {\bibinfo {volume} {20}},\ \bibinfo {pages} {9640}
  (\bibinfo {year} {2012})}\BibitemShut {NoStop}%
\bibitem [{\citenamefont {Christensen}\ \emph {et~al.}(2013)\citenamefont
  {Christensen}, \citenamefont {McCusker}, \citenamefont {Altepeter},
  \citenamefont {Calkins}, \citenamefont {Gerrits}, \citenamefont {Lita},
  \citenamefont {Miller}, \citenamefont {Shalm}, \citenamefont {Zhang},
  \citenamefont {Nam}, \citenamefont {Brunner}, \citenamefont {Lim},
  \citenamefont {Gisin},\ and\ \citenamefont {Kwiat}}]{CMA:prl13}%
  \BibitemOpen
  \bibfield  {author} {\bibinfo {author} {\bibfnamefont {B.~G.}\ \bibnamefont
  {Christensen}}, \bibinfo {author} {\bibfnamefont {K.~T.}\ \bibnamefont
  {McCusker}}, \bibinfo {author} {\bibfnamefont {J.~B.}\ \bibnamefont
  {Altepeter}}, \bibinfo {author} {\bibfnamefont {B.}~\bibnamefont {Calkins}},
  \bibinfo {author} {\bibfnamefont {T.}~\bibnamefont {Gerrits}}, \bibinfo
  {author} {\bibfnamefont {A.~E.}\ \bibnamefont {Lita}}, \bibinfo {author}
  {\bibfnamefont {A.}~\bibnamefont {Miller}}, \bibinfo {author} {\bibfnamefont
  {L.~K.}\ \bibnamefont {Shalm}}, \bibinfo {author} {\bibfnamefont
  {Y.}~\bibnamefont {Zhang}}, \bibinfo {author} {\bibfnamefont {S.-W.}\
  \bibnamefont {Nam}}, \bibinfo {author} {\bibfnamefont {N.}~\bibnamefont
  {Brunner}}, \bibinfo {author} {\bibfnamefont {C.~C.~W.}\ \bibnamefont {Lim}},
  \bibinfo {author} {\bibfnamefont {N.}~\bibnamefont {Gisin}}, \ and\ \bibinfo
  {author} {\bibfnamefont {P.~G.}\ \bibnamefont {Kwiat}},\ }\href@noop {}
  {\bibfield  {journal} {\bibinfo  {journal} {Phys. Rev. Lett.}\ }\textbf
  {\bibinfo {volume} {111}},\ \bibinfo {pages} {130406} (\bibinfo {year}
  {2013})}\BibitemShut {NoStop}%
\bibitem [{\citenamefont {Boitier}\ \emph {et~al.}(2014)\citenamefont
  {Boitier}, \citenamefont {Orieux}, \citenamefont {Autebert}, \citenamefont
  {Lemaître}, \citenamefont {Galopin}, \citenamefont {Manquest}, \citenamefont
  {Sirtori}, \citenamefont {Favero}, \citenamefont {Leo},\ and\ \citenamefont
  {Ducci}}]{BOA:prl14}%
  \BibitemOpen
  \bibfield  {author} {\bibinfo {author} {\bibfnamefont {F.}~\bibnamefont
  {Boitier}}, \bibinfo {author} {\bibfnamefont {A.}~\bibnamefont {Orieux}},
  \bibinfo {author} {\bibfnamefont {C.}~\bibnamefont {Autebert}}, \bibinfo
  {author} {\bibfnamefont {A.}~\bibnamefont {Lemaître}}, \bibinfo {author}
  {\bibfnamefont {E.}~\bibnamefont {Galopin}}, \bibinfo {author} {\bibfnamefont
  {C.}~\bibnamefont {Manquest}}, \bibinfo {author} {\bibfnamefont
  {C.}~\bibnamefont {Sirtori}}, \bibinfo {author} {\bibfnamefont
  {I.}~\bibnamefont {Favero}}, \bibinfo {author} {\bibfnamefont
  {G.}~\bibnamefont {Leo}}, \ and\ \bibinfo {author} {\bibfnamefont
  {S.}~\bibnamefont {Ducci}},\ }\href@noop {} {\bibfield  {journal} {\bibinfo
  {journal} {Phys. Rev. Lett.}\ }\textbf {\bibinfo {volume} {112}},\ \bibinfo
  {pages} {183901} (\bibinfo {year} {2014})}\BibitemShut {NoStop}%
\bibitem [{\citenamefont {Zhu}\ \emph {et~al.}(2012)\citenamefont {Zhu},
  \citenamefont {Tang}, \citenamefont {Qian}, \citenamefont {Helt},
  \citenamefont {Liscidini}, \citenamefont {Sipe}, \citenamefont {Corbari},
  \citenamefont {Canagasabey}, \citenamefont {Ibsen},\ and\ \citenamefont
  {Kazansky}}]{ZTQ:prl12}%
  \BibitemOpen
  \bibfield  {author} {\bibinfo {author} {\bibfnamefont {E.~Y.}\ \bibnamefont
  {Zhu}}, \bibinfo {author} {\bibfnamefont {Z.}~\bibnamefont {Tang}}, \bibinfo
  {author} {\bibfnamefont {L.}~\bibnamefont {Qian}}, \bibinfo {author}
  {\bibfnamefont {L.~G.}\ \bibnamefont {Helt}}, \bibinfo {author}
  {\bibfnamefont {M.}~\bibnamefont {Liscidini}}, \bibinfo {author}
  {\bibfnamefont {J.~E.}\ \bibnamefont {Sipe}}, \bibinfo {author}
  {\bibfnamefont {C.}~\bibnamefont {Corbari}}, \bibinfo {author} {\bibfnamefont
  {A.}~\bibnamefont {Canagasabey}}, \bibinfo {author} {\bibfnamefont
  {M.}~\bibnamefont {Ibsen}}, \ and\ \bibinfo {author} {\bibfnamefont {P.~G.}\
  \bibnamefont {Kazansky}},\ }\href@noop {} {\bibfield  {journal} {\bibinfo
  {journal} {Phys. Rev. Lett.}\ }\textbf {\bibinfo {volume} {108}},\ \bibinfo
  {pages} {213902} (\bibinfo {year} {2012})}\BibitemShut {NoStop}%
\bibitem [{\citenamefont {Lee}\ \emph {et~al.}(2006)\citenamefont {Lee},
  \citenamefont {Chen}, \citenamefont {Liang}, \citenamefont {Li},
  \citenamefont {Voss},\ and\ \citenamefont {Kumar}}]{LCL:opex06}%
  \BibitemOpen
  \bibfield  {author} {\bibinfo {author} {\bibfnamefont {K.~F.}\ \bibnamefont
  {Lee}}, \bibinfo {author} {\bibfnamefont {J.}~\bibnamefont {Chen}}, \bibinfo
  {author} {\bibfnamefont {C.}~\bibnamefont {Liang}}, \bibinfo {author}
  {\bibfnamefont {X.}~\bibnamefont {Li}}, \bibinfo {author} {\bibfnamefont
  {P.~L.}\ \bibnamefont {Voss}}, \ and\ \bibinfo {author} {\bibfnamefont
  {P.}~\bibnamefont {Kumar}},\ }\href@noop {} {\bibfield  {journal} {\bibinfo
  {journal} {Opt. Lett.}\ }\textbf {\bibinfo {volume} {31}},\ \bibinfo {pages}
  {1905} (\bibinfo {year} {2006})}\BibitemShut {NoStop}%
\bibitem [{\citenamefont {Bell}\ \emph {et~al.}(2012)\citenamefont {Bell},
  \citenamefont {Clark}, \citenamefont {Tame}, \citenamefont {Halder},
  \citenamefont {Fulconis}, \citenamefont {Wadsworth},\ and\ \citenamefont
  {Rarity}}]{BCT:njp12}%
  \BibitemOpen
  \bibfield  {author} {\bibinfo {author} {\bibfnamefont {B.}~\bibnamefont
  {Bell}}, \bibinfo {author} {\bibfnamefont {A.~S.}\ \bibnamefont {Clark}},
  \bibinfo {author} {\bibfnamefont {M.~S.}\ \bibnamefont {Tame}}, \bibinfo
  {author} {\bibfnamefont {M.}~\bibnamefont {Halder}}, \bibinfo {author}
  {\bibfnamefont {J.}~\bibnamefont {Fulconis}}, \bibinfo {author}
  {\bibfnamefont {W.}~\bibnamefont {Wadsworth}}, \ and\ \bibinfo {author}
  {\bibfnamefont {J.}~\bibnamefont {Rarity}},\ }\href@noop {} {\bibfield
  {journal} {\bibinfo  {journal} {New J. Phys.}\ }\textbf {\bibinfo {volume}
  {14}},\ \bibinfo {pages} {023021} (\bibinfo {year} {2012})}\BibitemShut
  {NoStop}%
\bibitem [{\citenamefont {Silverstone}\ \emph {et~al.}(2014)\citenamefont
  {Silverstone}, \citenamefont {Bonneau}, \citenamefont {Ohira}, \citenamefont
  {Suzuki}, \citenamefont {Yoshida}, \citenamefont {Iizuka}, \citenamefont
  {Ezaki}, \citenamefont {Natarajan}, \citenamefont {Tanner}, \citenamefont
  {Hadfield}, \citenamefont {Zwiller}, \citenamefont {Marshall}, \citenamefont
  {Rarity}, \citenamefont {O'Brien},\ and\ \citenamefont
  {Thompson}}]{SBO:natphoton14}%
  \BibitemOpen
  \bibfield  {author} {\bibinfo {author} {\bibfnamefont {J.~W.}\ \bibnamefont
  {Silverstone}}, \bibinfo {author} {\bibfnamefont {D.}~\bibnamefont
  {Bonneau}}, \bibinfo {author} {\bibfnamefont {K.}~\bibnamefont {Ohira}},
  \bibinfo {author} {\bibfnamefont {N.}~\bibnamefont {Suzuki}}, \bibinfo
  {author} {\bibfnamefont {H.}~\bibnamefont {Yoshida}}, \bibinfo {author}
  {\bibfnamefont {N.}~\bibnamefont {Iizuka}}, \bibinfo {author} {\bibfnamefont
  {M.}~\bibnamefont {Ezaki}}, \bibinfo {author} {\bibfnamefont {C.~M.}\
  \bibnamefont {Natarajan}}, \bibinfo {author} {\bibfnamefont {M.~G.}\
  \bibnamefont {Tanner}}, \bibinfo {author} {\bibfnamefont {R.~H.}\
  \bibnamefont {Hadfield}}, \bibinfo {author} {\bibfnamefont {V.}~\bibnamefont
  {Zwiller}}, \bibinfo {author} {\bibfnamefont {G.~D.}\ \bibnamefont
  {Marshall}}, \bibinfo {author} {\bibfnamefont {J.~G.}\ \bibnamefont
  {Rarity}}, \bibinfo {author} {\bibfnamefont {J.~L.}\ \bibnamefont {O'Brien}},
  \ and\ \bibinfo {author} {\bibfnamefont {M.~G.}\ \bibnamefont {Thompson}},\
  }\href@noop {} {\bibfield  {journal} {\bibinfo  {journal} {Nature Photon.}\
  }\textbf {\bibinfo {volume} {8}},\ \bibinfo {pages} {104} (\bibinfo {year}
  {2014})}\BibitemShut {NoStop}%
\bibitem [{\citenamefont {Huber}\ \emph {et~al.}(2014)\citenamefont {Huber},
  \citenamefont {Predojevic}, \citenamefont {Khoshnegar}, \citenamefont
  {Dalacu}, \citenamefont {Poole}, \citenamefont {Majedi},\ and\ \citenamefont
  {Weihs}}]{HPK:nanolett14}%
  \BibitemOpen
  \bibfield  {author} {\bibinfo {author} {\bibfnamefont {T.}~\bibnamefont
  {Huber}}, \bibinfo {author} {\bibfnamefont {A.}~\bibnamefont {Predojevic}},
  \bibinfo {author} {\bibfnamefont {M.}~\bibnamefont {Khoshnegar}}, \bibinfo
  {author} {\bibfnamefont {D.}~\bibnamefont {Dalacu}}, \bibinfo {author}
  {\bibfnamefont {P.~J.}\ \bibnamefont {Poole}}, \bibinfo {author}
  {\bibfnamefont {H.}~\bibnamefont {Majedi}}, \ and\ \bibinfo {author}
  {\bibfnamefont {G.}~\bibnamefont {Weihs}},\ }\href@noop {} {\bibfield
  {journal} {\bibinfo  {journal} {Nano Lett.}\ }\textbf {\bibinfo {volume}
  {14}},\ \bibinfo {pages} {7107} (\bibinfo {year} {2014})}\BibitemShut
  {NoStop}%
\bibitem [{\citenamefont {Versteegh}\ \emph {et~al.}(2014)\citenamefont
  {Versteegh}, \citenamefont {Reimer}, \citenamefont {J\"ons}, \citenamefont
  {Dalacu}, \citenamefont {Poole}, \citenamefont {Gulinatti}, \citenamefont
  {Giudice},\ and\ \citenamefont {Zwiller}}]{VRJ:natcomm14}%
  \BibitemOpen
  \bibfield  {author} {\bibinfo {author} {\bibfnamefont {M.~A.~M.}\
  \bibnamefont {Versteegh}}, \bibinfo {author} {\bibfnamefont {M.~E.}\
  \bibnamefont {Reimer}}, \bibinfo {author} {\bibfnamefont {K.~D.}\
  \bibnamefont {J\"ons}}, \bibinfo {author} {\bibfnamefont {D.}~\bibnamefont
  {Dalacu}}, \bibinfo {author} {\bibfnamefont {P.~J.}\ \bibnamefont {Poole}},
  \bibinfo {author} {\bibfnamefont {A.}~\bibnamefont {Gulinatti}}, \bibinfo
  {author} {\bibfnamefont {A.}~\bibnamefont {Giudice}}, \ and\ \bibinfo
  {author} {\bibfnamefont {V.}~\bibnamefont {Zwiller}},\ }\href@noop {}
  {\bibfield  {journal} {\bibinfo  {journal} {Nature Commun.}\ }\textbf
  {\bibinfo {volume} {5}},\ \bibinfo {pages} {5298} (\bibinfo {year}
  {2014})}\BibitemShut {NoStop}%
\bibitem [{\citenamefont {Trotta}\ \emph {et~al.}(2015)\citenamefont {Trotta},
  \citenamefont {Martin-Sanchez}, \citenamefont {Wildmann}, \citenamefont
  {Piredda}, \citenamefont {Reindl}, \citenamefont {Schimpf}, \citenamefont
  {Zallo}, \citenamefont {Schmidt}, \citenamefont {Stroj}, \citenamefont
  {Edlinger},\ and\ \citenamefont {Rastelli}}]{TMW:arxiv15}%
  \BibitemOpen
  \bibfield  {author} {\bibinfo {author} {\bibfnamefont {R.}~\bibnamefont
  {Trotta}}, \bibinfo {author} {\bibfnamefont {J.}~\bibnamefont
  {Martin-Sanchez}}, \bibinfo {author} {\bibfnamefont {J.~S.}\ \bibnamefont
  {Wildmann}}, \bibinfo {author} {\bibfnamefont {G.}~\bibnamefont {Piredda}},
  \bibinfo {author} {\bibfnamefont {M.}~\bibnamefont {Reindl}}, \bibinfo
  {author} {\bibfnamefont {C.}~\bibnamefont {Schimpf}}, \bibinfo {author}
  {\bibfnamefont {E.}~\bibnamefont {Zallo}}, \bibinfo {author} {\bibfnamefont
  {O.~G.}\ \bibnamefont {Schmidt}}, \bibinfo {author} {\bibfnamefont
  {S.}~\bibnamefont {Stroj}}, \bibinfo {author} {\bibfnamefont
  {J.}~\bibnamefont {Edlinger}}, \ and\ \bibinfo {author} {\bibfnamefont
  {A.}~\bibnamefont {Rastelli}},\ }\href@noop {} {\bibfield  {journal}
  {\bibinfo  {journal} {ArXiv preprint arXiv:1507.06612}\ } (\bibinfo {year}
  {2015})}\BibitemShut {NoStop}%
\bibitem [{\citenamefont {Fedrizzi}\ \emph {et~al.}(2009)\citenamefont
  {Fedrizzi}, \citenamefont {Ursin}, \citenamefont {Herbst}, \citenamefont
  {Nespoli}, \citenamefont {Prevedel}, \citenamefont {Scheidl}, \citenamefont
  {Tiefenbacher}, \citenamefont {Jennewein},\ and\ \citenamefont
  {Zeilinger}}]{FUH:natphys09}%
  \BibitemOpen
  \bibfield  {author} {\bibinfo {author} {\bibfnamefont {A.}~\bibnamefont
  {Fedrizzi}}, \bibinfo {author} {\bibfnamefont {R.}~\bibnamefont {Ursin}},
  \bibinfo {author} {\bibfnamefont {T.}~\bibnamefont {Herbst}}, \bibinfo
  {author} {\bibfnamefont {M.}~\bibnamefont {Nespoli}}, \bibinfo {author}
  {\bibfnamefont {R.}~\bibnamefont {Prevedel}}, \bibinfo {author}
  {\bibfnamefont {T.}~\bibnamefont {Scheidl}}, \bibinfo {author} {\bibfnamefont
  {F.}~\bibnamefont {Tiefenbacher}}, \bibinfo {author} {\bibfnamefont
  {T.}~\bibnamefont {Jennewein}}, \ and\ \bibinfo {author} {\bibfnamefont
  {A.}~\bibnamefont {Zeilinger}},\ }\href@noop {} {\bibfield  {journal}
  {\bibinfo  {journal} {Nature Phys.}\ }\textbf {\bibinfo {volume} {5}},\
  \bibinfo {pages} {389} (\bibinfo {year} {2009})}\BibitemShut {NoStop}%
\bibitem [{\citenamefont {Yin}\ \emph {et~al.}(2012)\citenamefont {Yin},
  \citenamefont {Ren}, \citenamefont {Lu}, \citenamefont {Cao}, \citenamefont
  {Yong}, \citenamefont {Wu}, \citenamefont {Liu}, \citenamefont {Liao},
  \citenamefont {Zhou}, \citenamefont {Jiang}, \citenamefont {Cai},
  \citenamefont {Xu}, \citenamefont {Pan}, \citenamefont {Jia}, \citenamefont
  {Huang}, \citenamefont {Yin}, \citenamefont {Wang}, \citenamefont {Chen},
  \citenamefont {Peng},\ and\ \citenamefont {Pan}}]{YRL:nature12}%
  \BibitemOpen
  \bibfield  {author} {\bibinfo {author} {\bibfnamefont {J.}~\bibnamefont
  {Yin}}, \bibinfo {author} {\bibfnamefont {J.~G.}\ \bibnamefont {Ren}},
  \bibinfo {author} {\bibfnamefont {H.}~\bibnamefont {Lu}}, \bibinfo {author}
  {\bibfnamefont {Y.}~\bibnamefont {Cao}}, \bibinfo {author} {\bibfnamefont
  {H.~L.}\ \bibnamefont {Yong}}, \bibinfo {author} {\bibfnamefont {Y.~P.}\
  \bibnamefont {Wu}}, \bibinfo {author} {\bibfnamefont {C.}~\bibnamefont
  {Liu}}, \bibinfo {author} {\bibfnamefont {S.~K.}\ \bibnamefont {Liao}},
  \bibinfo {author} {\bibfnamefont {F.}~\bibnamefont {Zhou}}, \bibinfo {author}
  {\bibfnamefont {Y.}~\bibnamefont {Jiang}}, \bibinfo {author} {\bibfnamefont
  {X.~D.}\ \bibnamefont {Cai}}, \bibinfo {author} {\bibfnamefont
  {P.}~\bibnamefont {Xu}}, \bibinfo {author} {\bibfnamefont {G.~S.}\
  \bibnamefont {Pan}}, \bibinfo {author} {\bibfnamefont {J.~J.}\ \bibnamefont
  {Jia}}, \bibinfo {author} {\bibfnamefont {Y.~M.}\ \bibnamefont {Huang}},
  \bibinfo {author} {\bibfnamefont {H.}~\bibnamefont {Yin}}, \bibinfo {author}
  {\bibfnamefont {J.~Y.}\ \bibnamefont {Wang}}, \bibinfo {author}
  {\bibfnamefont {Y.~A.}\ \bibnamefont {Chen}}, \bibinfo {author}
  {\bibfnamefont {C.~Z.}\ \bibnamefont {Peng}}, \ and\ \bibinfo {author}
  {\bibfnamefont {J.-W.}\ \bibnamefont {Pan}},\ }\href@noop {} {\bibfield
  {journal} {\bibinfo  {journal} {Nature}\ }\textbf {\bibinfo {volume} {488}},\
  \bibinfo {pages} {185} (\bibinfo {year} {2012})}\BibitemShut {NoStop}%
\bibitem [{\citenamefont {Fedrizzi}\ \emph {et~al.}(2007)\citenamefont
  {Fedrizzi}, \citenamefont {Herbst}, \citenamefont {Poppe}, \citenamefont
  {Jennewein},\ and\ \citenamefont {Zeilinger}}]{FHP:opex07}%
  \BibitemOpen
  \bibfield  {author} {\bibinfo {author} {\bibfnamefont {A.}~\bibnamefont
  {Fedrizzi}}, \bibinfo {author} {\bibfnamefont {T.}~\bibnamefont {Herbst}},
  \bibinfo {author} {\bibfnamefont {A.}~\bibnamefont {Poppe}}, \bibinfo
  {author} {\bibfnamefont {T.}~\bibnamefont {Jennewein}}, \ and\ \bibinfo
  {author} {\bibfnamefont {A.}~\bibnamefont {Zeilinger}},\ }\href@noop {}
  {\bibfield  {journal} {\bibinfo  {journal} {Opt. Express}\ }\textbf {\bibinfo
  {volume} {15}},\ \bibinfo {pages} {15377} (\bibinfo {year}
  {2007})}\BibitemShut {NoStop}%
\bibitem [{\citenamefont {Bao}\ \emph {et~al.}(2008)\citenamefont {Bao},
  \citenamefont {Qian}, \citenamefont {Yang}, \citenamefont {Zhang},
  \citenamefont {Chen}, \citenamefont {Yang},\ and\ \citenamefont
  {Pan}}]{BQY:prl08}%
  \BibitemOpen
  \bibfield  {author} {\bibinfo {author} {\bibfnamefont {X.~H.}\ \bibnamefont
  {Bao}}, \bibinfo {author} {\bibfnamefont {Y.}~\bibnamefont {Qian}}, \bibinfo
  {author} {\bibfnamefont {J.}~\bibnamefont {Yang}}, \bibinfo {author}
  {\bibfnamefont {H.}~\bibnamefont {Zhang}}, \bibinfo {author} {\bibfnamefont
  {Z.~B.}\ \bibnamefont {Chen}}, \bibinfo {author} {\bibfnamefont
  {T.}~\bibnamefont {Yang}}, \ and\ \bibinfo {author} {\bibfnamefont {J.-W.}\
  \bibnamefont {Pan}},\ }\href@noop {} {\bibfield  {journal} {\bibinfo
  {journal} {Phys. Rev. Lett.}\ }\textbf {\bibinfo {volume} {101}},\ \bibinfo
  {pages} {190501} (\bibinfo {year} {2008})}\BibitemShut {NoStop}%
\bibitem [{\citenamefont {Lim}\ \emph {et~al.}(2008)\citenamefont {Lim},
  \citenamefont {Yoshizawa}, \citenamefont {Tsuchida},\ and\ \citenamefont
  {Kikuchi}}]{LYT:opex08b}%
  \BibitemOpen
  \bibfield  {author} {\bibinfo {author} {\bibfnamefont {H.~C.}\ \bibnamefont
  {Lim}}, \bibinfo {author} {\bibfnamefont {A.}~\bibnamefont {Yoshizawa}},
  \bibinfo {author} {\bibfnamefont {H.}~\bibnamefont {Tsuchida}}, \ and\
  \bibinfo {author} {\bibfnamefont {K.}~\bibnamefont {Kikuchi}},\ }\href@noop
  {} {\bibfield  {journal} {\bibinfo  {journal} {Opt. Express}\ }\textbf
  {\bibinfo {volume} {16}},\ \bibinfo {pages} {22099} (\bibinfo {year}
  {2008})}\BibitemShut {NoStop}%
\bibitem [{\citenamefont {Zhou}\ \emph {et~al.}(2013)\citenamefont {Zhou},
  \citenamefont {Jiang}, \citenamefont {Ding}, \citenamefont {Shi},\ and\
  \citenamefont {Guo}}]{ZJD:pra13}%
  \BibitemOpen
  \bibfield  {author} {\bibinfo {author} {\bibfnamefont {Z.~Y.}\ \bibnamefont
  {Zhou}}, \bibinfo {author} {\bibfnamefont {Y.~K.}\ \bibnamefont {Jiang}},
  \bibinfo {author} {\bibfnamefont {D.~S.}\ \bibnamefont {Ding}}, \bibinfo
  {author} {\bibfnamefont {B.~S.}\ \bibnamefont {Shi}}, \ and\ \bibinfo
  {author} {\bibfnamefont {G.~C.}\ \bibnamefont {Guo}},\ }\href@noop {}
  {\bibfield  {journal} {\bibinfo  {journal} {Phys. Rev. A}\ }\textbf {\bibinfo
  {volume} {87}},\ \bibinfo {pages} {045806} (\bibinfo {year}
  {2013})}\BibitemShut {NoStop}%
\bibitem [{\citenamefont {Herbauts}\ \emph {et~al.}(2013)\citenamefont
  {Herbauts}, \citenamefont {Blauensteiner}, \citenamefont {Poppe},
  \citenamefont {Jennewein},\ and\ \citenamefont {H\"ubel}}]{HBP:opex13}%
  \BibitemOpen
  \bibfield  {author} {\bibinfo {author} {\bibfnamefont {I.}~\bibnamefont
  {Herbauts}}, \bibinfo {author} {\bibfnamefont {B.}~\bibnamefont
  {Blauensteiner}}, \bibinfo {author} {\bibfnamefont {A.}~\bibnamefont
  {Poppe}}, \bibinfo {author} {\bibfnamefont {T.}~\bibnamefont {Jennewein}}, \
  and\ \bibinfo {author} {\bibfnamefont {H.}~\bibnamefont {H\"ubel}},\
  }\href@noop {} {\bibfield  {journal} {\bibinfo  {journal} {Opt. Express}\
  }\textbf {\bibinfo {volume} {21}},\ \bibinfo {pages} {29013} (\bibinfo {year}
  {2013})}\BibitemShut {NoStop}%
\bibitem [{\citenamefont {Arahira}\ and\ \citenamefont
  {Murai}(2013)}]{AM:opex13}%
  \BibitemOpen
  \bibfield  {author} {\bibinfo {author} {\bibfnamefont {S.}~\bibnamefont
  {Arahira}}\ and\ \bibinfo {author} {\bibfnamefont {H.}~\bibnamefont
  {Murai}},\ }\href@noop {} {\bibfield  {journal} {\bibinfo  {journal} {Opt.
  Express}\ }\textbf {\bibinfo {volume} {21}},\ \bibinfo {pages} {7841}
  (\bibinfo {year} {2013})}\BibitemShut {NoStop}%
\bibitem [{\citenamefont {Zhu}\ \emph {et~al.}(2015)\citenamefont {Zhu},
  \citenamefont {Corbari}, \citenamefont {Gladyshev}, \citenamefont {Kazansky},
  \citenamefont {Lo},\ and\ \citenamefont {Qian}}]{ZCG:arxiv15}%
  \BibitemOpen
  \bibfield  {author} {\bibinfo {author} {\bibfnamefont {E.~Y.}\ \bibnamefont
  {Zhu}}, \bibinfo {author} {\bibfnamefont {C.}~\bibnamefont {Corbari}},
  \bibinfo {author} {\bibfnamefont {E.~V.}\ \bibnamefont {Gladyshev}}, \bibinfo
  {author} {\bibfnamefont {P.~G.}\ \bibnamefont {Kazansky}}, \bibinfo {author}
  {\bibfnamefont {H.-K.}\ \bibnamefont {Lo}}, \ and\ \bibinfo {author}
  {\bibfnamefont {L.}~\bibnamefont {Qian}},\ }\href@noop {} {\bibfield
  {journal} {\bibinfo  {journal} {ArXiv preprint arXiv:1506.03896}\ } (\bibinfo
  {year} {2015})}\BibitemShut {NoStop}%
\bibitem [{\citenamefont {Matsuda}\ \emph {et~al.}(2014)\citenamefont
  {Matsuda}, \citenamefont {Karkus}, \citenamefont {Nishi}, \citenamefont
  {Tsuchizawa}, \citenamefont {Munro}, \citenamefont {Takesue},\ and\
  \citenamefont {Yamada}}]{MKN:opex14}%
  \BibitemOpen
  \bibfield  {author} {\bibinfo {author} {\bibfnamefont {N.}~\bibnamefont
  {Matsuda}}, \bibinfo {author} {\bibfnamefont {P.}~\bibnamefont {Karkus}},
  \bibinfo {author} {\bibfnamefont {H.}~\bibnamefont {Nishi}}, \bibinfo
  {author} {\bibfnamefont {T.}~\bibnamefont {Tsuchizawa}}, \bibinfo {author}
  {\bibfnamefont {W.~J.}\ \bibnamefont {Munro}}, \bibinfo {author}
  {\bibfnamefont {H.}~\bibnamefont {Takesue}}, \ and\ \bibinfo {author}
  {\bibfnamefont {K.}~\bibnamefont {Yamada}},\ }\href@noop {} {\bibfield
  {journal} {\bibinfo  {journal} {Opt. Express}\ }\textbf {\bibinfo {volume}
  {22}},\ \bibinfo {pages} {22831} (\bibinfo {year} {2014})}\BibitemShut
  {NoStop}%
\bibitem [{\citenamefont {Ciurana}\ \emph {et~al.}(2014)\citenamefont
  {Ciurana}, \citenamefont {Martinez-Mateo}, \citenamefont {Peev},
  \citenamefont {Poppe}, \citenamefont {Walenta}, \citenamefont {Zbinden},\
  and\ \citenamefont {Martin}}]{CMP:opex14}%
  \BibitemOpen
  \bibfield  {author} {\bibinfo {author} {\bibfnamefont {A.}~\bibnamefont
  {Ciurana}}, \bibinfo {author} {\bibfnamefont {J.}~\bibnamefont
  {Martinez-Mateo}}, \bibinfo {author} {\bibfnamefont {M.}~\bibnamefont
  {Peev}}, \bibinfo {author} {\bibfnamefont {A.}~\bibnamefont {Poppe}},
  \bibinfo {author} {\bibfnamefont {N.}~\bibnamefont {Walenta}}, \bibinfo
  {author} {\bibfnamefont {H.}~\bibnamefont {Zbinden}}, \ and\ \bibinfo
  {author} {\bibfnamefont {V.}~\bibnamefont {Martin}},\ }\href@noop {}
  {\bibfield  {journal} {\bibinfo  {journal} {Opt. Express}\ }\textbf {\bibinfo
  {volume} {22}},\ \bibinfo {pages} {1576} (\bibinfo {year}
  {2014})}\BibitemShut {NoStop}%
\bibitem [{\citenamefont {Ghalbouni}\ \emph {et~al.}(2013)\citenamefont
  {Ghalbouni}, \citenamefont {Agha}, \citenamefont {Frey}, \citenamefont
  {Diamanti},\ and\ \citenamefont {Zaquine}}]{GAF:ol13}%
  \BibitemOpen
  \bibfield  {author} {\bibinfo {author} {\bibfnamefont {J.}~\bibnamefont
  {Ghalbouni}}, \bibinfo {author} {\bibfnamefont {I.}~\bibnamefont {Agha}},
  \bibinfo {author} {\bibfnamefont {R.}~\bibnamefont {Frey}}, \bibinfo {author}
  {\bibfnamefont {E.}~\bibnamefont {Diamanti}}, \ and\ \bibinfo {author}
  {\bibfnamefont {I.}~\bibnamefont {Zaquine}},\ }\href@noop {} {\bibfield
  {journal} {\bibinfo  {journal} {Opt. Lett.}\ }\textbf {\bibinfo {volume}
  {38}},\ \bibinfo {pages} {34} (\bibinfo {year} {2013})}\BibitemShut {NoStop}%
\bibitem [{\citenamefont {Clauser}\ \emph {et~al.}(1969)\citenamefont
  {Clauser}, \citenamefont {Horne}, \citenamefont {Shimony},\ and\
  \citenamefont {Holt}}]{CHSH}%
  \BibitemOpen
  \bibfield  {author} {\bibinfo {author} {\bibfnamefont {J.~F.}\ \bibnamefont
  {Clauser}}, \bibinfo {author} {\bibfnamefont {M.~A.}\ \bibnamefont {Horne}},
  \bibinfo {author} {\bibfnamefont {A.}~\bibnamefont {Shimony}}, \ and\
  \bibinfo {author} {\bibfnamefont {R.~A.}\ \bibnamefont {Holt}},\ }\href@noop
  {} {\bibfield  {journal} {\bibinfo  {journal} {Phys. Rev. Lett.}\ }\textbf
  {\bibinfo {volume} {23}},\ \bibinfo {pages} {880} (\bibinfo {year}
  {1969})}\BibitemShut {NoStop}%
\bibitem [{\citenamefont {Smirr}\ \emph {et~al.}(2013)\citenamefont {Smirr},
  \citenamefont {Deconinck}, \citenamefont {Frey}, \citenamefont {Agha},
  \citenamefont {Diamanti},\ and\ \citenamefont {Zaquine}}]{SDF:josab13}%
  \BibitemOpen
  \bibfield  {author} {\bibinfo {author} {\bibfnamefont {J.-L.}\ \bibnamefont
  {Smirr}}, \bibinfo {author} {\bibfnamefont {M.}~\bibnamefont {Deconinck}},
  \bibinfo {author} {\bibfnamefont {R.}~\bibnamefont {Frey}}, \bibinfo {author}
  {\bibfnamefont {I.}~\bibnamefont {Agha}}, \bibinfo {author} {\bibfnamefont
  {E.}~\bibnamefont {Diamanti}}, \ and\ \bibinfo {author} {\bibfnamefont
  {I.}~\bibnamefont {Zaquine}},\ }\href@noop {} {\bibfield  {journal} {\bibinfo
   {journal} {J. Opt. Soc. Am. B}\ }\textbf {\bibinfo {volume} {30}},\ \bibinfo
  {pages} {288} (\bibinfo {year} {2013})}\BibitemShut {NoStop}%
\bibitem [{\citenamefont {Si}\ and\ \citenamefont {Cheng}(2003)}]{CC:book03}%
  \BibitemOpen
  \bibfield  {author} {\bibinfo {author} {\bibfnamefont {Y.~C.}\ \bibnamefont
  {Si}}\ and\ \bibinfo {author} {\bibfnamefont {Y.}~\bibnamefont {Cheng}},\
  }\href@noop {} {\emph {\bibinfo {title} {WDM Technologies, Passive optical
  components}}},\ edited by\ \bibinfo {editor} {\bibfnamefont {A.~K.}\
  \bibnamefont {Dutta}}, \bibinfo {editor} {\bibfnamefont {N.~K.}\ \bibnamefont
  {Dutta}}, \ and\ \bibinfo {editor} {\bibfnamefont {M.}~\bibnamefont
  {Fujiwara}}\ (\bibinfo  {publisher} {Academic Press, 2003},\ \bibinfo {year}
  {2003})\ pp.\ \bibinfo {pages} {39--78}\BibitemShut {NoStop}%
\bibitem [{\citenamefont {Kwiat}\ \emph {et~al.}(1994)\citenamefont {Kwiat},
  \citenamefont {Steinberg}, \citenamefont {Chiao}, \citenamefont {Eberhard},\
  and\ \citenamefont {Petroff}}]{KSC:ao94}%
  \BibitemOpen
  \bibfield  {author} {\bibinfo {author} {\bibfnamefont {P.~G.}\ \bibnamefont
  {Kwiat}}, \bibinfo {author} {\bibfnamefont {A.~M.}\ \bibnamefont
  {Steinberg}}, \bibinfo {author} {\bibfnamefont {R.~Y.}\ \bibnamefont
  {Chiao}}, \bibinfo {author} {\bibfnamefont {P.~H.}\ \bibnamefont {Eberhard}},
  \ and\ \bibinfo {author} {\bibfnamefont {M.~D.}\ \bibnamefont {Petroff}},\
  }\href@noop {} {\bibfield  {journal} {\bibinfo  {journal} {Appl.Opt.}\
  }\textbf {\bibinfo {volume} {33}},\ \bibinfo {pages} {1844} (\bibinfo {year}
  {1994})}\BibitemShut {NoStop}%
\bibitem [{\citenamefont {White}\ \emph {et~al.}(2001)\citenamefont {White},
  \citenamefont {James}, \citenamefont {Munro},\ and\ \citenamefont
  {Kwiat}}]{WJM:pra01}%
  \BibitemOpen
  \bibfield  {author} {\bibinfo {author} {\bibfnamefont {A.~G.}\ \bibnamefont
  {White}}, \bibinfo {author} {\bibfnamefont {D.~F.~V.}\ \bibnamefont {James}},
  \bibinfo {author} {\bibfnamefont {W.~J.}\ \bibnamefont {Munro}}, \ and\
  \bibinfo {author} {\bibfnamefont {P.~G.}\ \bibnamefont {Kwiat}},\ }\href@noop
  {} {\bibfield  {journal} {\bibinfo  {journal} {Phys. Rev. A}\ }\textbf
  {\bibinfo {volume} {65}},\ \bibinfo {pages} {012301} (\bibinfo {year}
  {2001})}\BibitemShut {NoStop}%
\end{thebibliography}%


\section*{appendix: Calculation of the demultiplexer figures of merit}

\subsection{Quality factor $\zeta_{Q}$.}\label{appendix:zetaQ}
In this section we derive the expression for the quality factor $\zeta_{Q}$ taking into account loss, spectral shape of the channel transmission and symmetry of the signal and idler channels.

The maximum visibility that can be obtained is given by the expression $V_{\text{max}} = 1/(1+2P_{AC}/P_{TC})$, where $P_{AC}$ and $P_{TC}$ are the probabilities of measuring accidental and true coincidences, respectively. The brightness is proportional to $P_{TC}$.
Under the assumption that the only limitation to the photon pair correlations is due to multiple pair generation, and that the global transmission of the whole channel is low,
\begin{eqnarray}\label{eq:P}
P_{i}=p_0I_{1_i}T_{i}K_T\nonumber \\
P_{AC}\simeq p_0^2I_{1_A}I_{1_B}T_{A}T_{B}K_T^2G_T^2\\
P_{TC}\simeq p_0I_{2}T_{A}T_{B}K_TG_T\nonumber
\end{eqnarray}
where $p_0$ is the pair generation probability density, $T_{i}$, the maximum transmission of the channel $i$, $K_T$ is the total detection time per second, and $G_T$ the size of the coincidence gate. Replacing $P_{AC}$ and $P_{TC}$ by their expression in Eq.\ref{eq:P}, we can derive the maximum possible visibility:
\begin{equation}
V_{\text{max}} = \frac{1}{1+2p_0I_{1_A}I_{1_B}K_TG_T/I_{2}}
\end{equation}
In this model, when $p_0$ increases, the number of true coincidences increases linearly but the visibility decreases because of the quadratic increase of the number of accidental coincidences.
If $P_{AC}/P_{TC} \ll 1$, the decrease is almost linear with increasing $p_0$, or with increasing brightness. The most useful demultiplexer will be the one with the slowest decrease of the maximum visibility with respect to the brightness. The quality factor of the demultiplexer can therefore be defined as proportional to the inverse of the slope :
\begin{equation}
\zeta_{Q}= \frac{P_{AC}}{P_{TC}^2}=\frac{I_{2}^2T_{A}T_{B}}{I_{1_A}I_{1_B}}\\
\end{equation}

\subsection{Quality factor $\eta$.}\label{appendix:PMD}
Here we model the effect of noise, losses and polarization mode dispersion (PMD) on the density matrix of the two-photon state.

Starting from a maximally entangled state $\rho_{1} = \ketbrad{\Phi^+}$, noise and losses are captured by the visibility $V_0$ and the density matrix becomes the following Werner state:
\begin{eqnarray}
\rho_{2} &=& V_0\rho_1+\frac{1-V_0}{4}\one\nonumber\\
 &=&
\begin{pmatrix}
\frac{1}{2}V_0+\frac{1-V_0}{4}& 0 & 0 & \frac{1}{2}V_0\\
0 & \frac{1-V_0}{4} & 0 & 0\\
0 & 0 & \frac{1-V_0}{4} & 0\\
\frac{1}{2}V_0 & 0 & 0 & \frac{1}{2}V_0+\frac{1-V_0}{4}
\end{pmatrix}
.
\end{eqnarray}
The effect of PMD is then modeled through a dephasing channel acting on one of the photons:
\begin{eqnarray}
\rho_{3} &=& \frac{1+\eta}{2}\rho_2+\frac{1-\eta}{2}(\one\otimes\sigma_z)\rho_2(\one\otimes\sigma_z)^\dagger\nonumber\\
 &=&
\begin{pmatrix}
\frac{1}{2}V_0+\frac{1-V_0}{4}& 0 & 0 & \frac{1}{2}\eta V_0\\
0 & \frac{1-V_0}{4} & 0 & 0\\
0 & 0 & \frac{1-V_0}{4} & 0\\
\frac{1}{2}\eta V_0 & 0 & 0 & \frac{1}{2}V_0+\frac{1-V_0}{4},
\end{pmatrix}
\end{eqnarray}
where $\eta = \int{f_s(t)f_i(\tau_{PMD}-t)dt}$ is the temporal overlap between the vertical photon wavepackets of the signal and idler demultiplexer channels, assuming that the horizontal ones overlap perfectly, $\tau_{\text{PMD}}$ is the delay due to the differential PMD between the two channels and $f_s(t)$ and $f_i(t)$ are the normalized temporal transmission functions of the signal and idler channels.

The visibility in the diagonal basis is then given by: $V_{45} = \eta V_0$, and the Bell parameter is expressed as:
\begin{equation}
S = \sqrt{2}(V_0+V_{45}) = \sqrt{2}V_0(1+\eta).
\end{equation}

\end{document}